\documentclass[letterpaper,twocolumn,conference]{IEEEtran}
\usepackage[T1]{fontenc}
\usepackage{color}
\usepackage{float}
\usepackage{calc}
\usepackage{units}
\usepackage{amsmath}
\usepackage{graphicx}
\usepackage{nomencl}
\providecommand{\printnomenclature}{\printglossary}
\providecommand{\makenomenclature}{\makeglossary}
\makenomenclature
\usepackage[unicode=true,
 bookmarks=true,bookmarksnumbered=true,bookmarksopen=true,bookmarksopenlevel=1,
 breaklinks=false,pdfborder={0 0 0},pdfborderstyle={},backref=false,colorlinks=false]
 {hyperref}
\hypersetup{pdftitle={Drone Based 3D},
 pdfauthor={Marc Gauger, Stephan ten Brink},
 pdfpagelayout=OneColumn, pdfnewwindow=true, pdfstartview=XYZ, plainpages=false, draft}

\makeatletter

\pdfpageheight\paperheight
\pdfpagewidth\paperwidth

\floatstyle{ruled}
\newfloat{algorithm}{tbp}{loa}
\providecommand{\algorithmname}{Algorithm}
\floatname{algorithm}{\protect\algorithmname}

\usepackage{cite}
\usepackage[super]{nth}
\usepackage{amsmath,amsthm,array}

\usepackage{tikz}
\usetikzlibrary{shapes.geometric}
\usetikzlibrary{decorations.pathreplacing}
\usetikzlibrary{patterns}
\usetikzlibrary{arrows}
\usetikzlibrary{3d}
\usetikzlibrary{decorations}
\usetikzlibrary{calc, positioning}
\usetikzlibrary{arrows}
\usetikzlibrary{fit}
\usetikzlibrary{angles, quotes, backgrounds,decorations.markings,shapes,shapes.multipart,positioning}
\usetikzlibrary{matrix}

\usepackage{xcolor}

\usepackage{pgfplots}

\pgfplotsset{compat=newest}
\usepgfplotslibrary{fillbetween}
\usetikzlibrary{intersections}
\usetikzlibrary{calc}

\definecolor{uniSblue}{RGB}{0,65,145}
\definecolor{uniSlightblue}{RGB}{0,190,255}
\definecolor{uniSgray}{RGB}{62, 68, 76}
\definecolor{uniSyellow}{RGB}{255, 213, 0}
\definecolor{uniAnthrazit}{HTML}{323232}

\definecolor{uniSred}{RGB}{230, 0, 50}
\definecolor{uniSgreen}{RGB}{0, 200, 50}

\usepackage{eso-pic}
\usepackage{pdfpages}
\usepackage{esint}
\usepackage{stackengine}

\usepackage{pgfplots}
\usepgfplotslibrary{external} 
\usepgfplotslibrary{polar}

\usepackage{algorithm,algpseudocode}
\usepackage{amssymb}
\usepackage{romannum}

\usepackage{cleveref}
\crefname{figure}{Fig.}{Figures}
\crefname{section}{Section}{Sections}
\crefname{subsection}{Paragraph}{Paragraphs}
\crefname{table}{Table}{Tables}
\crefname{equation}{}{}

\usepackage{caption}

\usepackage{tcolorbox}
\usepackage{adjustbox}

\usepackage{xfrac}
\usepackage{siunitx}
\DeclareSIUnit\bit{b}
\DeclareSIUnit\bpcu{bpcu}
    \usepackage{pdfcomment}
    \usepackage{environ}
    \RenewEnviron{comment}{\color{magenta}{\BODY}}

\usepackage[paper=letterpaper, headheight=2in,headsep=0.7in, top=0.8in,left=0.7in,right=0.7in,bottom=0.8in]{geometry}
\usepackage[printonlyused,withpage]{acronym}

\renewcommand*{\nomenclature}[3][]{\ac{#2}}
 
\usepackage{algorithm,algpseudocode}

\tikzset{>=latex}
\usetikzlibrary{intersections}

\newacro{svd}[SVD]{singular value decomposition} 
\newacro{qd}[QD]{quasi-determinisctic} 
\newacro{fcc}[FCC]{federal communications commission} 
\newacro{wlan}[WLAN]{wireless local area network} 
\newacro{sls}[SLS]{sector level sweep} 
\newacro{brp}[BRP]{beam refinement phase} 
\newacro{paa}[PAA]{phased array antenna}
\newacro{hmimo}[H-MIMO]{hybrid-MIMO}
\newacro{siso}[SISO]{single-input single-output}
\newacro{ofdm}[OFDM]{orthogonal frequency division multiplex}
\newacro{rssi}[RSSI]{received signal strength indicator}

\newacro{cir}[CIR]{channel impulse response}
\newacro{dft}[DFT]{discrete Fourier transform}
\newacro{cm}[CM]{channel model}
\newacro{11ad}[11ad]{IEEE~802.11ad}
\newacro{11ay}[11ay]{IEEE~802.11ay}
\newacro{11az}[11az]{IEEE~802.11az}
\newacro{sta}[STA]{station}
\newacro{ap}[AP]{access point}
\newacro{cr}[CR]{conference room}
\newacro{rf}[RF]{radio frequency}
\newacro{pdf}[PDF]{probability density function}
\newacro{gps}[GPS]{global positioning system}
\newacro{ftm}[FTM]{Fine Time Measurement}
\newacro{dft}[DFT]{discrete Fourier transform}
\newacro{doa}[DOA]{direction of arrival}
\newacro{dod}[DOD]{direction of departure}
\newacro{aoa}[AOA]{angle of arrival}
\newacro{aod}[AOD]{angle of departure}

\newacro{nlos}[NLOS]{non-\ac{los}}
\newacro{nsoe}[NSOE]{nonlinear system of equations}
\newacro{ae}[AE]{angular estimation}
\newacro{rx}[RX]{receiver}
\newacro{tx}[TX]{transmitter}
\newacro{ura}[URA]{uniform rectangular array}
\newacro{awv}[AWV]{antenna weight vector}
\newacro{tof}[TOF]{time of flight}
\newacro{toa}[TOA]{time of arrival}
\newacro{aoad}[AOA/D]{\ac{aoa}/\ac{aod}}
\newacro{lm}[LM]{Levenberg-Marquardt}
\newacro{roi}[ROI]{region of interest}
\newacro{mac}[MAC]{medium access conrol}
\newacro{phy}[PHY]{physical layer}
\newacro{qd}[QD]{quasi deterministic}
\newacro{2d}[2D]{two-dimensional}
\newacro{3d}[PHY]{three-dimensional}
\newacro{ccdf}[CCDF]{complementary cumulative distribution function}
\newacro{cdf}[CDF]{cumulative distribution function}
\newacro{ura}[URA]{uniform rectangular array}

\newacro{rtk}[RTK]{real time kinematic}
\newacro{c1}[conf. \Romannum{1}]{configuration \Romannum{1}}
\newacro{c2}[conf. \Romannum{2}]{configuration \Romannum{2}}
\newacro{c3}[conf. \Romannum{3}]{configuration \Romannum{3}}

\newacro{gps}[GPS]{global positioning system}
\newacro{ofdm}[OFDM]{orthogonal frequency-division multiplexing}
\newacro{mimo}[MIMO]{multiple-input multiple-output}
\newacro{mrt}[MRT]{Maximum Ratio Transmission}
\newacro{po}[PO]{Phase Only}
\newacro{fpga}[FPGA]{Field Programmable Gate Array}
\newacro{dc}[DC]{direct current}
\newacro{gps-rtk}[GPS-RTK]{Global Positioning System - Real Time Kinematic}
\newacro{cfo}[CFO]{carrier frequency offset}
\newacro{bpsk}[BPSK]{binary phase shift keying}
\newacro{rf}[RF]{radio frequency}
\newacro{ssd}[SSD]{Solid-State-Drive}
\newacro{csi}[CSI]{channel state information}
\newacro{adc}[A/D converter]{analog to digital converter}
\newacro{pll}[PLL]{phase locked loop}
\newacro{snr}[SNR]{signal-to-noise ratio}
\newacro{sir}[SIR]{signal-to-interference ratio}
\newacro{rx}[RX]{receiver}
\newacro{pep}[PEP]{peak envelope power}
\newacro{eirp}[EIRP]{effective (or equivalent) isotropic radiated power}
\newacro{los}[LOS]{line-of-sight}
\newacro{cfo}[CFO]{carrier frequency offset}

\makeatother

\begin{document}
\title{Massive MIMO Channel Measurements and Achievable Rates in a Residential
Area}
\author{}
\author{\vspace{-2.5cm}\author{\IEEEauthorblockA{Marc Gauger\IEEEauthorrefmark{1},  Maximilian Arnold\IEEEauthorrefmark{1} and Stephan ten Brink\IEEEauthorrefmark{1}\\
\IEEEauthorrefmark{1}Institute of Telecommunications, Pfaffenwaldring 47, University of Stuttgart, 70569 Stuttgart, Germany}\\
\vspace*{-1.5cm}}}
\maketitle
\begin{abstract}
In this paper we present a measurement set-up for massive MIMO channel
sounding that shows very good long-term phase stability. Initial measurements
were performed in a residential area to evaluate different conventional
precoding schemes such as maximum ratio transmission and phase only
precoding. A massive amount of data points was collected, with 924
times 64 complex channel weights per data point. Each data point is
position-tagged using differential GPS with real-time kinematik, achieving
better than 35cm position accuracy in more than 90\% of the collected
data points, making this dataset a rich resource for, e.g., further
studying machine learning based, data-driven approaches in wireless
communications.

\vspace*{0cm}\acresetall
\end{abstract}

\IEEEpeerreviewmaketitle{}

\section{Introduction\label{sec:Introduction}}

Massive \nomenclature{mimo}{ } is a key enabling technology for the
future wireless ``5G'' standard and beyond \cite{EmilBjoernson2017,MMIMO5GCommMag,HoydisTenBrinkMMIMO}.
To evaluate massive \nomenclature{mimo}{ } algorithms and achievable
sum-rate capacities, several channel models have been established
in the literature \cite{Jaeckel2014,7881048}. However, channel models
can only provide an abstract view considering the most important wireless
propagation phenomena, and do often model specific communication scenarios
and hardware impairments only rudimentarily. Thus, actual channel
measurements in typical coverage settings, like performed in this
paper, offer the potential of providing much more realistic estimates
on the actual achievable data rates and their particular distribution
over the spatial coverage region. 

\section{Measurement Set-Up}

The objective of this measurement campaign is to obtain actual measured
channel data (i.e., CSI, channel state information) of a typical residential
massive multiple input, multiple output (MIMO) antenna set-up. For
studying the effects of multiuser operation, position-labeled single-input/single
output (SIMO) measurements are required. Conceptually, the measurements
could be performed using two possible set-ups when assuming channel
reciprocity: (1) multiple transmit antennas at the basestation, and
one antenna at the mobile receiver, or, (2) multiple receive antennas
at the basestation, and one antenna at the mobile transmitter. Option
(1) requires a potentially large number of orthogonal pilots and perfect
frequency and time-alignment of the multiple transmitters, but simplifies
receiver post-processing; option (2) simplifies pilot design but requires
much more involved post-processing of the mulitple received signals.
We opted for set-up (2) as receiver imperfections, e.g., \nomenclature{cfo}{ },
can be more conveniently compensated on a per-antenna basis via post-processing;
also, the mobile transmitter having only a single antenna is
easier to implement and more lightweight to carry around.

\subsection{Portable Transmitter}

\begin{figure}[tbh]
\begin{centering}
\includegraphics{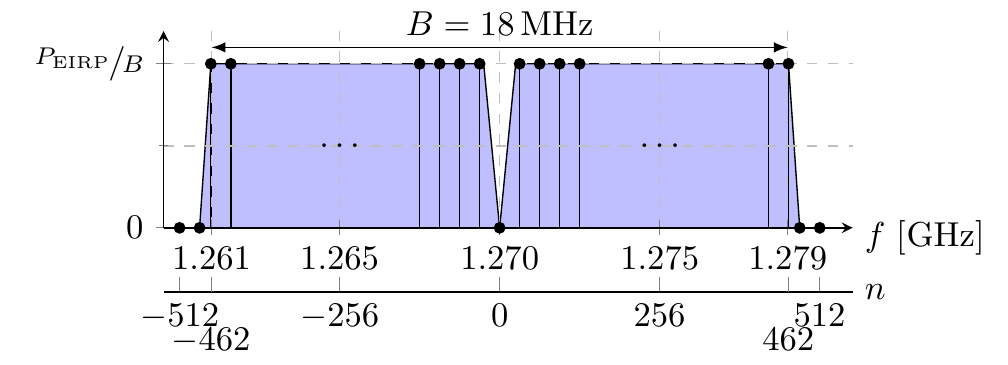}
\par\end{centering}
\caption{Spectrum of transmitted OFDM signal and corresponding subcarrier index
$n$\label{fig:Spectrum-of-transmitter}}
\end{figure}
The transmitter is based on an Ettus USRP B210 \cite{USRP}. Its \nomenclature{fpga}{  }
was programmed to generate \nomenclature{ofdm}{ } symbols of $B=\unit[18]{MHz}$
effective bandwidth. An \nomenclature{ofdm}{ } sample rate of $R_{s}=\unit[\frac{1}{T_{s}}=20]{MS/s}$
and a number of $N_{\text{sub}}=1024$ subcarriers are used. The subcarrier
spacing is about $\Delta f_{\text{sub}}=\frac{R_{s}}{N_{\text{sub}}}\approx\unit[20]{kHz}$.
At either band edge, 49-50 subcarriers were set to zero to relax the
requirements for analog filtering; also, the \nomenclature{dc}{ }
subcarrier was set to zero for carrier leakage reduction; thus, in
total 924 subcarriers are effectively used, as depicted in Fig. \ref{fig:Spectrum-of-transmitter}.
The carrier frequency was set to the unlicensed ham radio frequency
$\unit[1.27]{GHz}$ so that a rather large transmit \nomenclature{pep}{ }
of $\unit[18]{W}$ ($P_{\text{TX}}=\unit[42]{dBm}$) in combination with
a dipole antenna could be used. In this frequency range a \nomenclature{pep}{ }
of up to $\unit[750]{W}$ is allowed for licensed ham radio. The transmitter
dipole antenna has a gain of about $G_{\text{TX}}=\unit[9]{dBi}$ and is
vertically polarized, yielding an \nomenclature{eirp}{ } of about
$\unit[51]{dBm}$.

\begin{figure}[tbh]
\begin{centering}
\includegraphics[width=0.9\columnwidth]{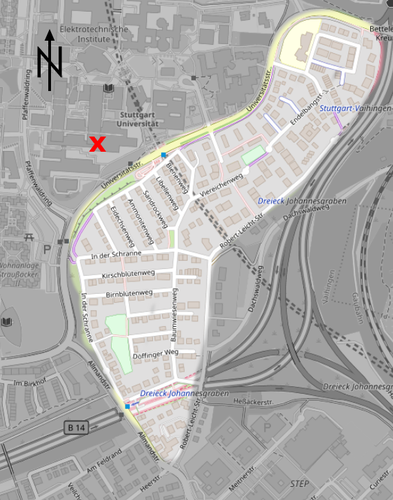}
\par\end{centering}
\caption{Map of considered residential area of dimension $\unit[600]{m}\times\unit[800]{m}$;
basestation location marked with a \textcolor{red}{red ``X''}\label{fig:Residential-area}}
\end{figure}

\begin{figure}[tbh]
\begin{centering}
\begin{center}
\tikzset{>=latex}

\begin{tikzpicture}[scale=0.4]
\definecolor{darkgreen}{rgb}{0.0, 0.2, 0.13};
\draw[thick] (-0.25,0) -- (20,0);
\draw[thick,->] (0,-13.5) -- (0,5.25);
\draw (0,6.5) node [] {Power};
\draw (0,5.75) node [] {dBm};
\draw (-0.5,0) node [] {0};

\draw[red] (-0.9,4) node [] {$P_{\text{TX}}$};
\draw[dashed,very thick,red] (0,4.5) -- (20,4.5);
\draw[red] (6,5) node [] {$P_{\text{EIRP}} \approx$ 50 dBm};
\draw[thick,red] (0,4) -- (1,4);

\draw[thick,red] (0.9,4) -- (0.9,4.5);
\draw[red] (1.5,5) node [] {$G_{\text{TX}}$};
\draw[thick,red] (0.8,4.5) -- (2.2,4.5);

\draw[thick,blue,->] (2.1,4.5) -- (2.1,-10);
\draw[blue] (5.6,3.5) node [] {path loss -110 dB};
\draw[thick,blue] (2,-10) -- (3,-10);

\draw[very thick,darkgreen] (2.9,-10) -- (2.9,-9.5);
\draw[very thick,darkgreen] (3.45,-8.75) node [] {$G_{\text{RX}}$};
\draw[very thick,darkgreen] (2.8,-9.5) -- (4.2,-9.5);

\draw[very thick,darkgreen] (4.1,-9.5) -- (4.1,-10.2);
\draw[very thick,darkgreen] (6.2,-9.8) node [] {cable loss};
\draw[very thick,darkgreen] (3.8,-10.2) -- (8.5,-10.2);

\draw[very thick,darkgreen,->] (8.3,-10.2) -- (8.3,-5.7);
\draw[very thick,darkgreen] (6.2,-7.8) node [] {amplifier};
\draw[very thick,darkgreen] (8,-5.7) -- (9.5,-5.7);

\draw[very thick,darkgreen] (9.4,-5.7) -- (9.4,-6.1);
\draw[very thick,darkgreen] (10.7,-5.2) node [] {noise figure (-1.6dB)};
\draw[very thick,darkgreen] (9,-6.1) -- (10.5,-6.1);

\draw[very thick,darkgreen] (9.8,-8.3) -- (11.8,-8.3);
\draw[very thick,darkgreen] (14.1,-7.3) node [] {multiplexer (-10 dB)};
\draw[very thick,darkgreen] (10.1,-8.3) -- (10.1,-6.1);

\draw[very thick,darkgreen] (11.3,-10.3) -- (11.3,-8.3);
\draw[very thick,darkgreen] (15.3,-9.4) node [] {summation (-10 dB)};
\draw[very thick,darkgreen] (11.1,-10.3) -- (17.2,-10.3);

\draw[dashed,very thick,orange] (0,-10.3) -- (20,-10.3);
\draw[orange] (-0.9,-9.7) node [] {$P_{\text{RX}}$};

\draw[very thick,orange] (17.1,-10.3) -- (17.1,-12.3);
\draw[very thick,orange] (14.05,-11.2) node [] {SNR = 30 dB};

\draw[very thick,magenta] (0,-12.3) -- (20,-12.3);
\draw[fill, magenta!20]  (0,-12.3) rectangle (20,-14.3);
\draw[magenta] (10,-12.9) node [] {(thermal and quantization noise in 20MHz)=-97dBm};

\draw[black] (14.5,2) node [] {coverage (1.27 GHz) $\approx$ 2.1 km};
\end{tikzpicture}
\par\end{center}
\par\end{centering}
\caption{Overview of link budget from TX to RX, resulting in an expected coverage
radius of about 2.1km\label{fig:Link-budget}}
\end{figure}
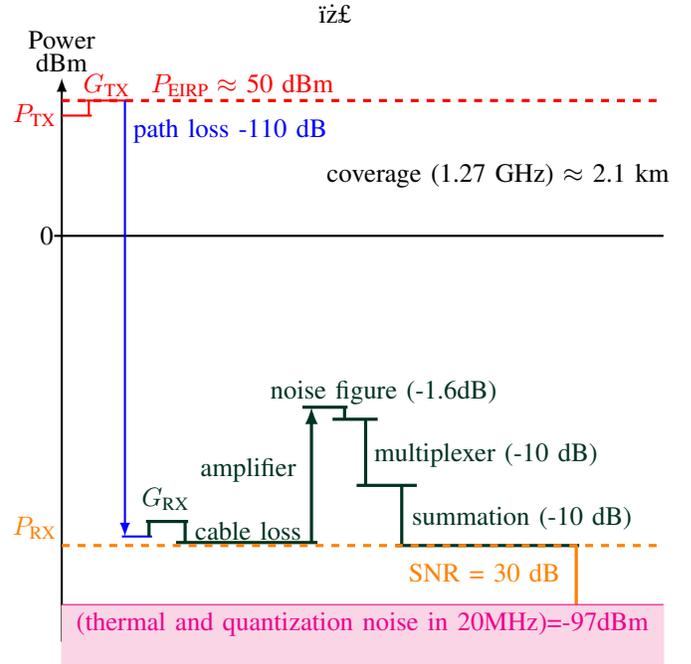

 Fig. \ref{fig:Link-budget} shows the link budget for a \nomenclature{los}{ }
channel. $G_{RX}$ is the \nomenclature{rx}{ } antenna gain. The
total measurement time was 8 hours, where the compact hand-wagon was
manually pushed around in a residential area of size about $\unit[600]{m}\times\unit[800]{m}$
(Fig. \ref{fig:Residential-area}).

\subsubsection{GPS positioning}

\begin{figure}[tbh]
\begin{centering}
\includegraphics[width=1\columnwidth]{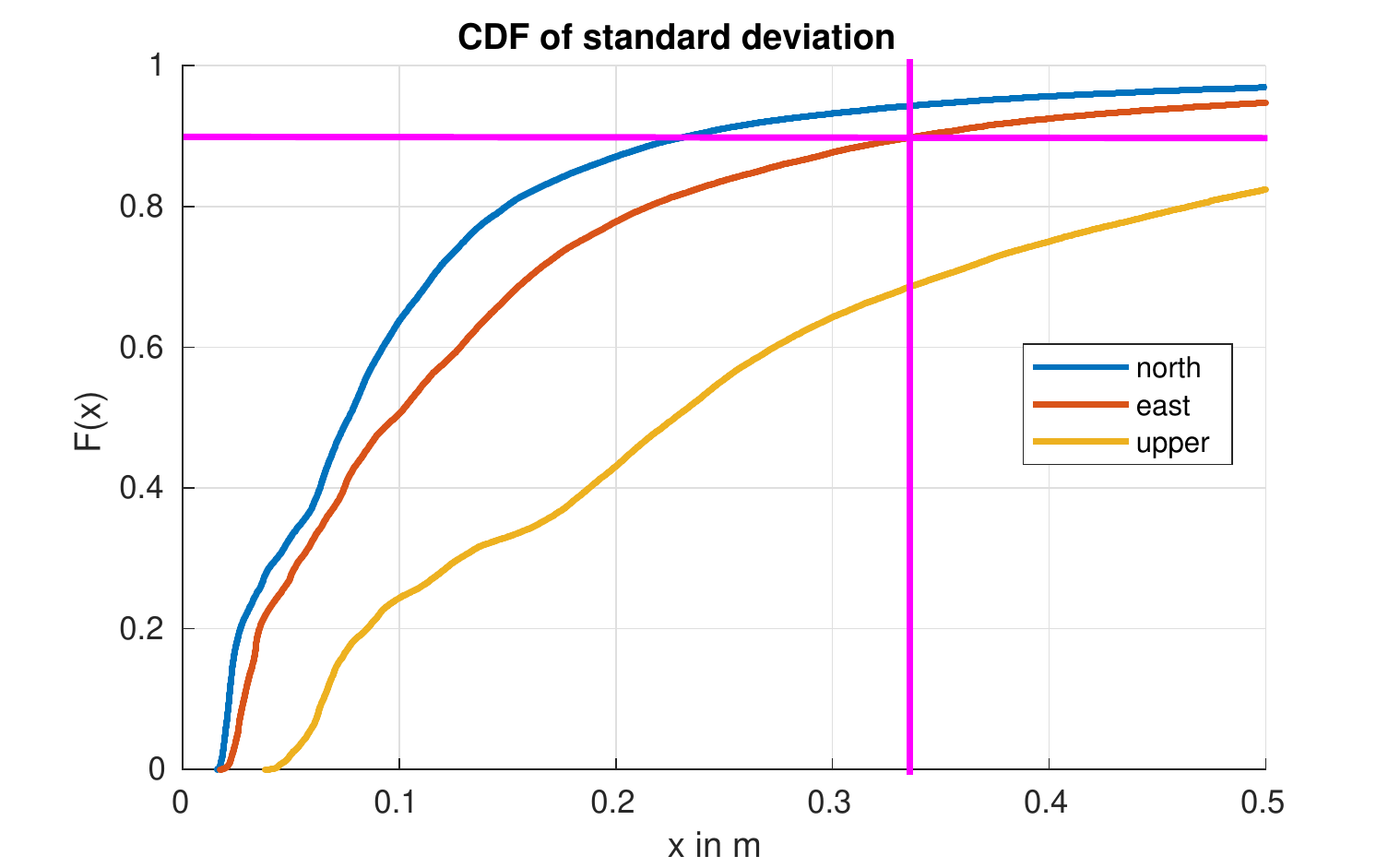}
\par\end{centering}
\caption{CDF of standard deviation of GPS position accuracy estimated by RTKLIB\cite{RTKLIB}\label{fig:GPS-Standard-Derivatio}}
\end{figure}

The position of the wagon was determined by the \nomenclature{gps-rtk}{ }
using RTKLIB \cite{RTKLIB}. The hardware used is based on two ``ublox''
NEO-M8T GPS receivers, with a dedicated GPS basestation fixed at the
roof-top of the Institute of Telecommunications, University of Stuttgart.
For illustration, Fig. \ref{fig:GPS-Standard-Derivatio} shows the
horizontal standard deviations with respect to North, East and the
vertical standard deviation with respect to ``Upper'' of the GPS
positioning of the RTKLIB. As can be seen (magenta lines in Fig. \ref{fig:GPS-Standard-Derivatio})
the $\left(x,y\right)-$accuracy was better than $\unit[34]{cm}$
in more than 90\% of the measurement points (with a $\unit[5]{Hz}$
update rate).

\subsubsection{OFDM frame structure}
\begin{figure}[tbh]
\centering{\begin{tikzpicture}[scale=0.8,transform shape]
\tikzset{>=latex}
\tikzstyle{box} = [draw,rounded corners=.1cm,minimum height=3em, minimum width=2em, align=center, thick]
\tikzstyle{vertical_box} = [draw,rounded corners=.1cm,minimum height=2em, minimum width=3em, align=center, thick,rotate=90]
\tikzstyle{data}=[rectangle split,rectangle split parts=2,draw,text centered]
\def\originx{0}
\def\originy{0}
\def\ueposx{0.4}
\def\ueposy{0.8}
\node[box,blue,minimum height=1cm,minimum width=1.5cm,fill=blue!10] (ue) at (\ueposx+0.42,\ueposy) {Pilot};
\node[box,blue,minimum height=1cm,minimum width=1.5cm,fill=blue!10] (ue) at (\ueposx+1.92,\ueposy) {Pilot};
\node[box,blue,minimum height=1cm,minimum width=1.5cm,fill=blue!10] (ue) at (\ueposx+4.98,\ueposy) {Pilot};
\node[box,red,minimum height=1cm,minimum width=1.5cm,fill=red!10] (ue) at (\ueposx+3.44,\ueposy) {Emb.\\ data};

\node [left] at (6.8,0.75) {$\hdots$};
\draw [thick, ->] (0,0) -- (0,2);
\draw [thick, ->] (0,0) -- (6,0);
\node [above] at (0,2) {$f$};
\node [right] at (6,0) {$t$};
\node [left] at (-0.1,1.3) {$f_{c}+\text{B}/2$};
\node [left] at (-0.1,0.3) {$f_{c}-\text{B}/2$};
\draw [thick, -] (-0.1,0.3) -- (0.1,0.3);
\draw [thick, -] (-0.1,1.3) -- (0.1,1.3);
\node at (0,-0.3) {$0$};
\draw [thick, -] (1.6,-0.1) -- (1.6,0.1);
\node at (1.6,-0.3) {$\unit[64]{\text{\textmu}s}$};
\draw [thick, -] (3.1,-0.1) -- (3.1,0.1);
\node at (3.1,-0.3) {$\unit[128]{\text{\textmu}s}$};
\draw [thick, -] (4.7,-0.1) -- (4.7,0.1);
\node at (4.7,-0.3) {$\unit[192]{\text{\textmu}s}$};
\end{tikzpicture}}\caption{Transmitted OFDM time-domain signal with pilot symbols and data symbols
carrying GPS data}
\end{figure}
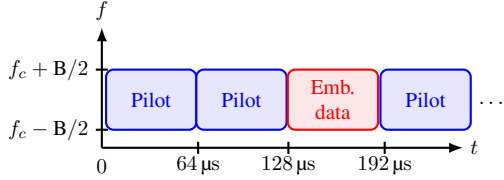
The \nomenclature{ofdm}{ } symbols are composed of 924 subcarriers.
The symbol duration with cyclic prefix is $T_{\text{OFDM}}=\unit[(N_{sub}-N_{\text{CP}})\cdot T_{s}=(1024+256)\cdot\frac{1}{\unit[20]{MS/s}}=64]{\text{\textmu}s}$,
where 25\% (i.e., $N_{\text{\text{CP}}}\cdot T_{s}=256\cdot\frac{1}{\unit[20]{MS/s}}=\unit[12.8]{\text{\textmu}s}$)
of the symbol duration was used as cyclic prefix (CP). One frame is
built of two pilot \nomenclature{ofdm}{ } symbols and one data \nomenclature{ofdm}{ }
symbol. The frames are repeated without any pause or null symbol.
The two pilot \nomenclature{ofdm}{ } symbols are used for channel
sounding and \nomenclature{cfo}{ } estimation. The data \nomenclature{ofdm}{ }
symbol is \nomenclature{bpsk}{ } modulated and contains an ID corresponding
to the actual GPS position. The bandwidth can be chosen freely, yet
needs to stay smaller than $\unit[56]{MHz}$ due to the limitations
of the USRP \cite{USRP}. For the channel measurements in this paper,
we used an $\unit[18]{MHz}$ \nomenclature{ofdm}{ } signal bandwidth
with sampling rate $\frac{1}{T_{s}}=\unit[20]{MS/s}$ at a carrier
frequency of $\unit[1.27]{GHz}$.

\subsection{Massive MIMO Receiver}

The receiver uses a 64-element antenna array with dual polarized patch
antennas, while only the vertical polarization was used. The 64 full \nomenclature{rf}{ }
chains are implemented in a multi-board/daughter-board configuration
to filter, amplify and downconvert (i.e., frequency shift) the respective
antenna signals. The 64 antenna signals of $\unit[18]{MHz}$ bandwidth,
all centered around the carrier frequency of $\unit[1.27]{GHz}$,
are shifted to different intermediate frequencies by means of separately
programmable downconverters so that the individual spectra do not
overlap in a spectral band from $\unit[10]{MHz}$ to $\unit[2]{GHz}$.
This composite ``frequency division multiple antenna''-signal is
then analog-to-digital converted and stored on a \nomenclature{ssd}{ }
drive by a digital oscilloscope Teledyne LeCroy WavePro 604HD, allowing
to measure spatial snapshots of 64 antennas times 924 subcarriers
complex \nomenclature{csi}{ } samples at a rate of \textasciitilde 11
measurements points per second (comp. \cite{Arnold2018} for more
details).

\begin{figure}[tbh]
\begin{centering}
\includegraphics[width=0.4\columnwidth]{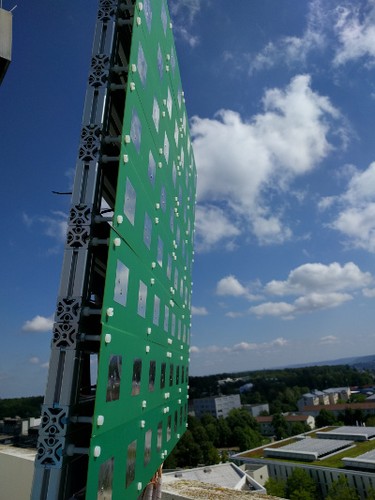}\hspace{1cm}\includegraphics[width=0.4\columnwidth]{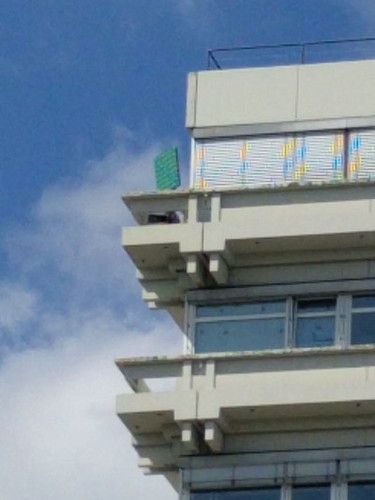}
\par\end{centering}
\caption{Photos of basestation receiver antenna with 64 antenna elements}
\end{figure}

\subsection{Extraction of Channel State Information\label{sec:Channel-Parameters}}

All 64 antenna signals are perfectly time synchronized in the digital
domain by jointly digitizing the composite ``frequency division multiple
antenna''-signal using the single \nomenclature{adc}{ } of the digital
scope (in fact, four \nomenclature{adc}{ }s of the scope were used
which, however, are perfectly synchronized). The repeated two pilots
symbols are used for the frame detection, channel estimation and for
 \nomenclature{cfo}{ } estimation by
employing 

\begin{equation}
\Delta f_{\text{CFO}}=\frac{1}{2\pi N_{\text{sub}}T_{s}}\sum_{\ell=0}^{N_{\text{sub}}-1}y\left(t_{0}+\ell\right)\cdot y\left(t_{0}+N_{\text{sub}}+N_{\text{CP}}+\ell\right)^{*}.\label{eq:CFO_Estimation}
\end{equation}

To obtain a measurement set-up which is stable over several hours,
all \nomenclature{csi}{ } phases are calibrated to a reference transmitter
which was at a fixed position only a few meters away from the antenna
array. The (narrowband) reference transmitter was set to a carrier
frequency of $\unit[1.257]{GHz}$ with a spectral bandwidth of $\unit[1.5]{MHz}$.
This way, any phase jitter/phase flips etc. of the \nomenclature{rf}{ }
chains that may occur due to the individual \nomenclature{pll}{ }
control loops are implicitly accounted for. Also, to account for different
gains of the \nomenclature{rf}{ } chains, the power of the individual
signals is further calibrated by estimating the \nomenclature{snr}{ }
per \nomenclature{rf}{ }-chain, and aligning the noise floor across
all antenna signals by respective amplitude scaling. To further reduce
the noise, i.e., to improve the \nomenclature{snr}{ }, the time-domain
channel impulse response was computed from the frequency domain channel
estimates \nomenclature{csi}{ }, and cut off so that multipath delays
up to $\unit[0.82]{ms}$ corresponding to a maximal path difference
of $\unit[1.92]{km}$ were accounted for (i.e., 128 samples at $T_{s}=\frac{1}{20}\mu s$).
The \nomenclature{rx}{ } antenna array was located at position $\left(x,y\right)=\left(0\mathrm{m},0\mathrm{m}\right)$
on a roof-top height of $\unit[40]{m}$ above ground and is marked
in the figures with a red cross (``X''). The \nomenclature{rx}{ }
antenna array was facing toward the South-East.

\begin{figure}[tbh]
\begin{centering}
\includegraphics[width=0.9\columnwidth]{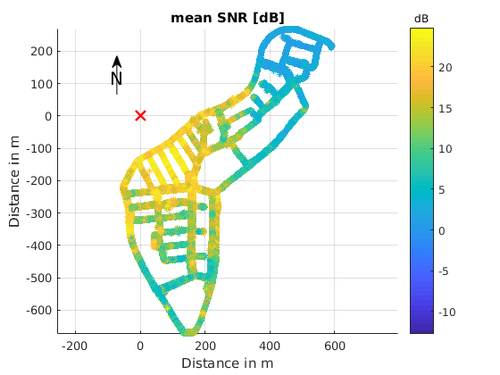}
\par\end{centering}
\caption{Plot of mean \nomenclature{snr}{ }\label{fig:mean-SNR} over coverage
area; basestation array antenna at $\left(x,y\right)=\left(0\mathrm{m},0\mathrm{m}\right)$,
40m height, facing South-East}
\end{figure}
Fig. \ref{fig:mean-SNR} plots the mean \nomenclature{snr}{ } computed
across all 64 \nomenclature{rx}{ } antennas. As expected, the area
in the North has to have a smaller \nomenclature{snr}{ } due to the
limited half power beam width of $\unit[69.1]{{^\circ}}$ of the receiver
array's patch antennas.

\begin{figure}[tbh]
\begin{centering}
\includegraphics[width=0.9\columnwidth]{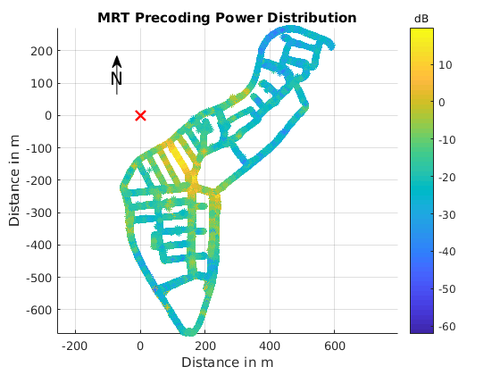}
\par\end{centering}
\caption{Illustration of the effect of \nomenclature{mrt}{ } precoding to
a point at $\left(x,y\right)=(110\mathrm{m},-120\mathrm{m})$\label{fig:MRT-Precoding-to}}
\end{figure}
As one sanity check for the measured data, Fig. \ref{fig:MRT-Precoding-to}
illustrates the distribution of the signal power if \nomenclature{mrt}{ }
precoding is applied to a position at $(110\mathrm{m},-120\mathrm{m})$,
resulting in a quite plausible beam/focus area around the desired
``target'' user point in comparison to the ``mean-SNR'' shown
in Fig. \ref{fig:mean-SNR} .

\section{K-Means Clustering: Phase Only vs MRT \label{sec:Clustering}}

One way of evaluating the achievable sum-rate of this residential
area setting is to apply a $k$-means clustering algorithm, to find
areas that are ``quasi'' mutually orthogonal to each other, i.e.,
the vector product $<X,Y>$ is very small, with $X$ taken from cluster
$i$, and $Y$ taken from cluster $j$, with $2\le i,j\le K$. An
overview of clustering schemes in massive \nomenclature{mimo}{ }
can be found in \cite{8385445}.

\subsection{Precoding and Clustering Algorithm}

Next, we outline the details of the clustering algorithm for the two
precoding techniques \nomenclature{mrt}{ } and \nomenclature{po}{ },
respectively. Algorithm \ref{alg:k-Means-MRT} describes a $k$-means
clustering with an \nomenclature{mrt}{ } precoding weight function,
and Algorithm \ref{alg:k-Means-PO} describes the version using a
\nomenclature{po}{ } weight function.

\begin{algorithm}[tbh]
\begin{algorithmic}[1]
\State{$M$ number of measurements}
\State{$k=1,...,K; \mathbf{C}\in\mathbb{C}(N_{\mathrm{antennas}} \times K)=\mathbf{c}_1,...,\mathbf{c}_K$}
\State{$\mathbf{H} \in \mathbb{C}(M \times N_{\mathrm{antennas}})$}
\State{Randomly choose K different indices of measurements}
\State{Define center of cluster as $\mathbf{c}_k = \mathbf{h}(k)$}
\For{Number of iterations}

\For{$k=1,...,K$}
\State{Find set $S_k$ of g with $\vert h(g) \mathbf{c}_{k}\vert = \text{max}_{\hat k} \vert\mathbf{H} \mathbf{c}_{\hat k}\vert$}
\State{ $\mathbf{c}_k=\frac{\sum_{S_k}h(g) } { \vert\sum_{S_k}h(g) \vert }$}
\EndFor
\EndFor
\end{algorithmic}

\caption{$k$-means MRT Algorithm\label{alg:k-Means-MRT}}
\end{algorithm}
\begin{algorithm}[tbh]
\begin{algorithmic}[1]
\State{$M$ number of measurements}
\State{$k=1,...,K; \mathbf{C}\in\mathbb{C}(N_{\mathrm{antennas}} \times K)=\mathbf{c}_1,...,\mathbf{c}_K$}
\State{$\mathbf{H} \in \mathbb{C}(M \times N_{\mathrm{antennas}})$}
\State{Randomly choose K different indices of measurements}
\State{Define center of cluster as $\mathbf{c}_k = e^{j\text{arg}(\mathbf{h}(k))}$}
\For{Number of iterations}

\For{$k=1,...,K$}
\State{Find set $S_k$ of g with $\vert h(g) \mathbf{c}_{k}\vert = \text{max}_{\hat k} \vert\mathbf{H} \mathbf{c}_{\hat k}\vert$}
\State{ $\mathbf{c}_k=\frac{\sum_{S_k}e^{j\text{arg}(h(g))} } { \sum_{S_k} 1  }$}
\EndFor
\EndFor
\end{algorithmic}

\caption{$k$-means PO Algorithm\label{alg:k-Means-PO}}
\end{algorithm}

\subsection{Results}

\begin{figure}[tbh]
\begin{centering}
\includegraphics[width=0.9\columnwidth]{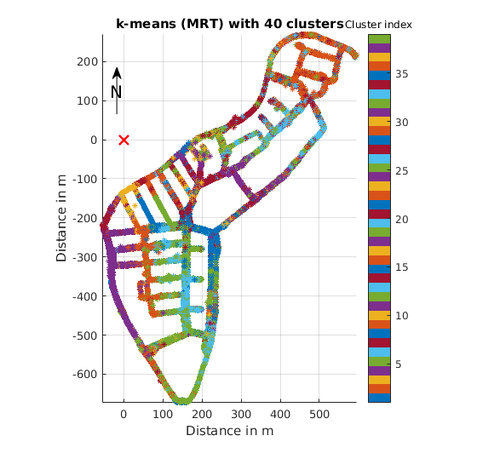}
\par\end{centering}
\caption{Result of a $k=40$ means \nomenclature{mrt}{ } clustering\label{fig:k=00003D40-Means-Maximum}}
\end{figure}

\begin{figure}[tbh]
\begin{centering}
\includegraphics[width=0.8\columnwidth]{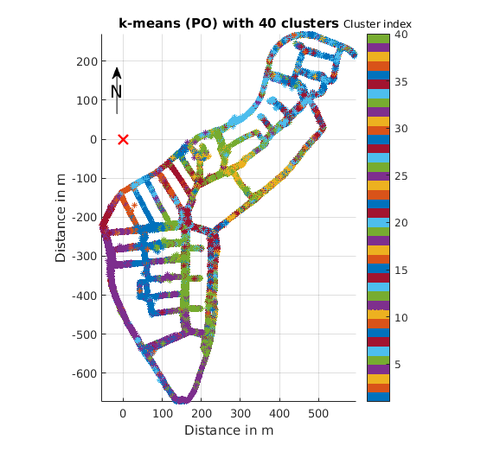}
\par\end{centering}
\caption{Result of a $k=40$ means \nomenclature{po}{ } clustering\label{fig:k=00003D40-Means-Phase}}
\end{figure}

Fig. \ref{fig:k=00003D40-Means-Maximum} shows exemplary the clustering
results for the $k$-means \nomenclature{mrt}{ }-based clustering
algorithm for $k=40$ clusters after 30 iterations. The positions
of the specific channel measurement are colored according to their
respective cluster. Fig. \ref{fig:k=00003D40-Means-Phase} shows the
clustering results for the $k$-means \nomenclature{po}{ } clustering
algorithm for the respective value of $k$, number of iterations and
the same randomly picked initial cluster centers. Obviously, channel
measurements which are locally close to each other are grouped into
the same cluster. The result shows an expected behavior from other
channel models. It is yet another sanity check for the measurements.
The regions with low \nomenclature{snr}{ } result in higher diversity
of clusters (with, yet, only small contribution to sum rate) and the
regions with higher \nomenclature{snr}{ }s show more separated clusters.
The reason for this effect is that two channels ``decorrelate''
with increasing noise power.

\begin{figure}[tbh]
\centering{}\includegraphics[width=0.9\columnwidth]{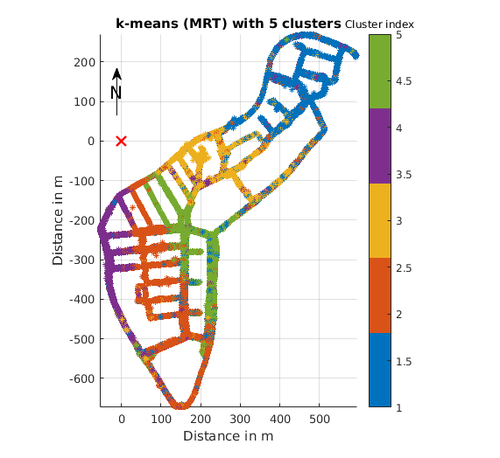}\caption{Result of a $k=5$ means \nomenclature{mrt}{ } clustering\label{fig:k5clusterMRT}}
\end{figure}
\begin{figure}[tbh]
\centering{}\includegraphics[width=0.8\columnwidth]{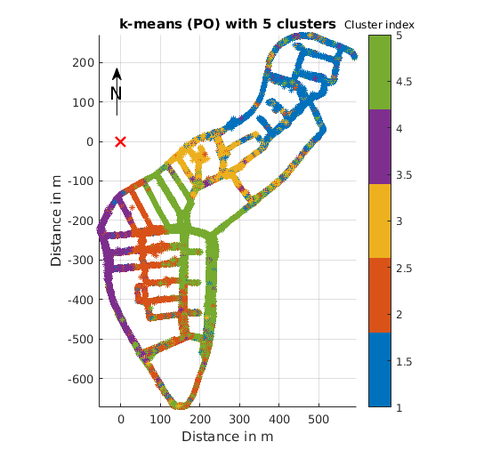}\caption{Result of a $k=5$ means \nomenclature{po}{ } clustering\label{fig:k5clusterPO}}
\end{figure}
\begin{figure}[tbh]
\begin{centering}
\includegraphics[width=1\columnwidth]{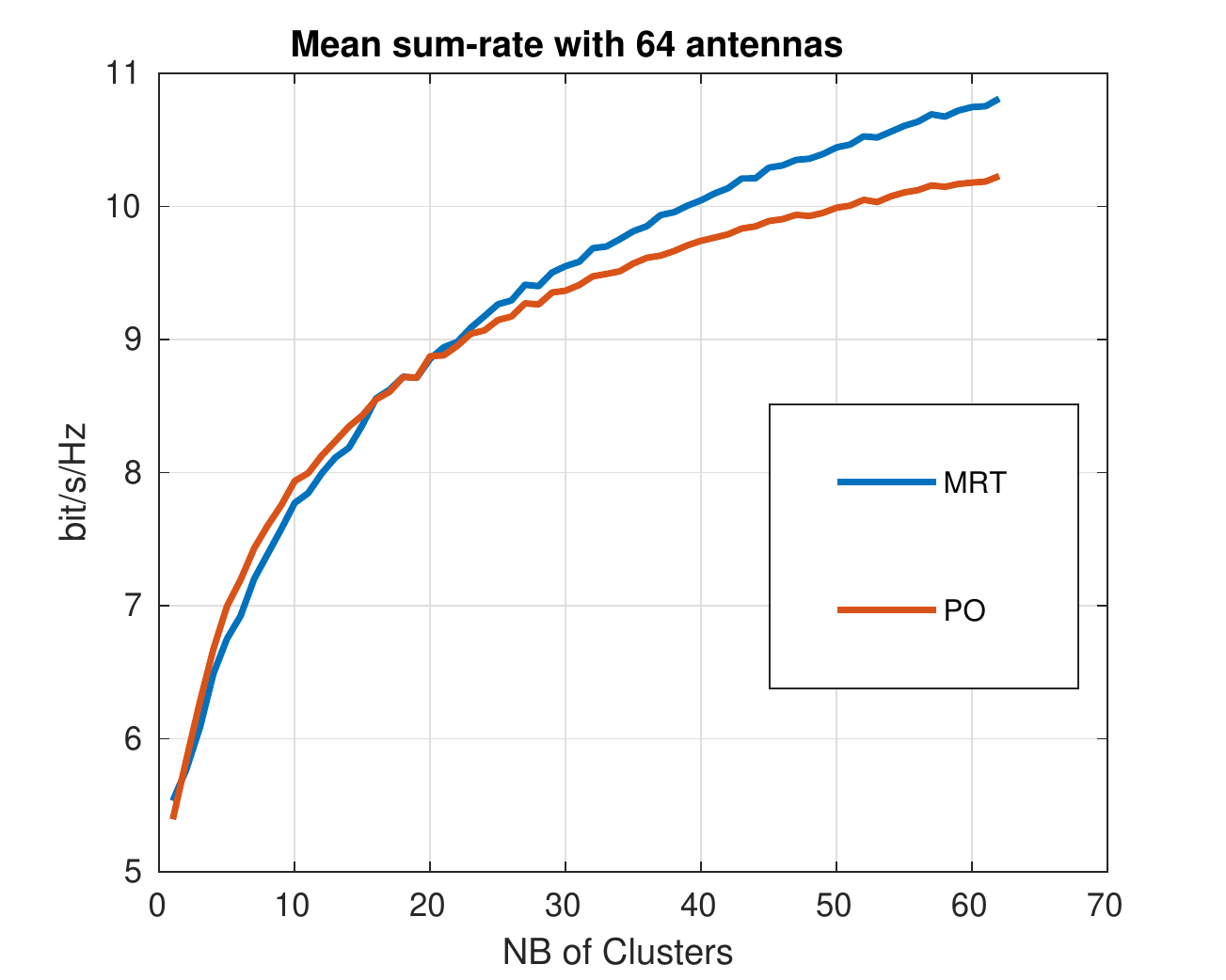}
\par\end{centering}
\caption{Mean sum-rate versus number of clusters $k$ (64 antennas) averaged
over 1000 cluster realizations on the measured data \label{fig:Median-Sum-Capacity}}
\end{figure}
\begin{figure}[tbh]
\begin{centering}
\includegraphics[width=1\columnwidth]{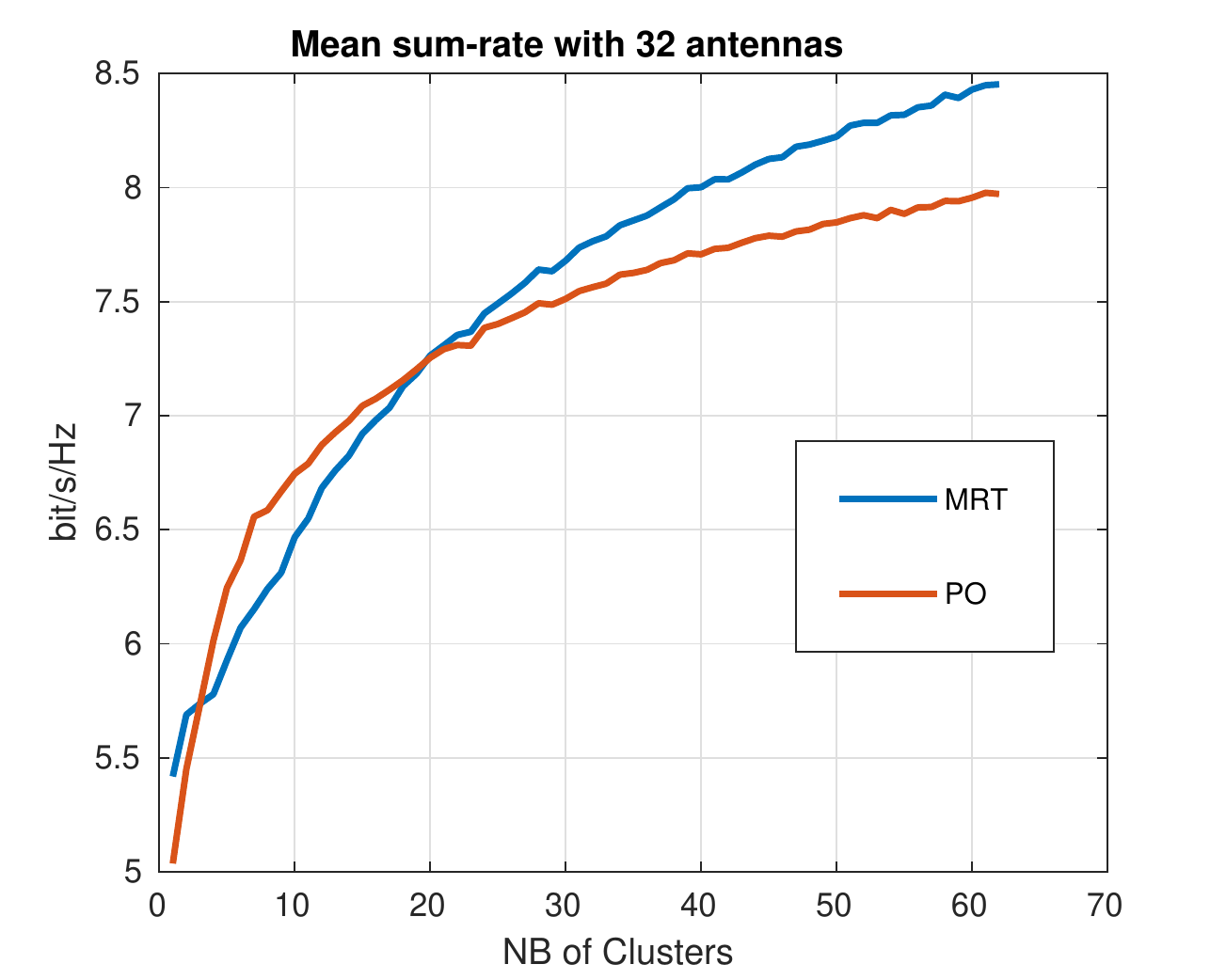}
\par\end{centering}
\caption{Mean sum-rate versus number of clusters $k$ (32 antennas) averaged
over 1000 cluster realizations on the measured data \label{fig:Median-Sum-Capacity-2}\label{fig:Median-Sum-Capacity-1}}
\end{figure}
To evaluate the results of the clustering algorithms, the sum-rate was
calculated while assuming an interference-limited channel. To account
for the noise, the \nomenclature{sir}{ } is clipped to a \nomenclature{sir}{ }
of $\unit[30]{dB}$.

The $\text{\text{SIR}}_{u}$ for user $u$ that belongs to cluster
$k$ is calculated according to
\begin{equation}
\mathrm{SIR_{u}}=\frac{\vert\mathrm{\boldsymbol{h}_{u}}\mathbf{c}_{k}\vert^{2}}{\sum_{\hat{k}\neq k}\vert\mathrm{\boldsymbol{h}}_{u}\mathrm{\boldsymbol{c}}_{\hat{k}}\vert^{2}}.\label{eq:SIR}
\end{equation}

The cluster $\text{\text{SIR}}_{k}$ is calculated by the median of
the $\text{\text{SIR}}_{u}$ of all users within this cluster $k$. The interference is calculated by the sum of the energy for user $u$ from all clusters $\hat{k}$ without cluster $k$. 

\begin{equation}
\text{SIR}_{k}=\text{median}(\text{SIR}_{u})\text{ with }\forall u\in\text{cluster k}\label{eq:SIR_cluster}
\end{equation}

The sum-rate $R_{\text{sum}}$ is calculated by the sum of the rate
of its respective cluster $\text{\text{SIR}}_{k}$.

\begin{equation}
R_{\text{sum}}=\sum_{k}\text{log}_{2}(1+\text{SIR}_{k})\label{eq:c_sum}
\end{equation}
Fig. \ref{fig:Median-Sum-Capacity} shows the mean of the sum-rate
after 1000 random initializations of the $k$-means clustering algorithm.
The increasing number of clusters results in a small increase of the
sum-rate. After $k\ge20$ clusters, \nomenclature{mrt}{ }-precoding
begins to significantly outperform \nomenclature{po}{ }-precoding.
Note that \nomenclature{mrt}{ }-precoding should be better than \nomenclature{po}{ }-precoding
for all $k$, indicating that the number of random initializations
for computing the median could still be increased.

However, the result also indicates that the lower complexity \nomenclature{po}{ }-precoding
only incurs a small loss with respect to \nomenclature{mrt}{ }-precoding
for lower number $k$ of clusters. For a small number of clusters,
e.g., $k=5$, both clustering algorithms converge to nearly the same
clusters. For small $k$, e.g., $k<20$, the \nomenclature{po}{ }-algorithm
approaches the \nomenclature{mrt}{ }-algorithm performance very closely.
Note that, the \nomenclature{po}{ }-algorithm considers only the
phase and is, thus, of much lower complexity (akin beam-forming).
The beam-forming characteristics can clearly be observed in Fig. \ref{fig:k5clusterPO}
and Fig. \ref{fig:k5clusterMRT}. If the number of clusters $k$ increases
to, e.g., $k=40$, the clusters have to be separated not only in the
angular direction but also in the radial distance, such that several
clusters are ``behind'' each other in radial direction; in this
case, the \nomenclature{mrt}{ }-algorithm outperforms the \nomenclature{po}{ }-based
cluster algorithm. Fig. \ref{fig:k=00003D40-Means-Maximum} and Fig.
\ref{fig:k=00003D40-Means-Phase} show the different clustering results
for the exact same initial cluster centers for $k=40$.

\section{Conclusions\label{sec:conclusions}}

In this paper, we presented massive \nomenclature{mimo}{ } channel
measurements, collecting a large quantity of channel state information
over a wide residential area. The $k$-means clustering algorithm
was applied to this \nomenclature{csi}{ }-data by using \nomenclature{mrt}{ }
and \nomenclature{po}{ }-precoding, respectively. As one interesting
result, \nomenclature{po}{ }-precoding, which is of much lower complexity,
turns out to lose only little in terms of sum-rate when compared to
the full \nomenclature{mrt}{ }-precoding.

\bibliographystyle{IEEEtran}
\bibliography{IEEEabrv,Schranne}

\settowidth{\nomlabelwidth}{gps-rtk}
\printnomenclature{}
\begin{IEEEbiography}[{\fbox{\begin{minipage}[t][1.25in]{1in}%
Replace this box by an image with a width of 1\,in and a height of
1.25\,in!%
\end{minipage}}}]{Your Name}
 All about you and the what your interests are.
\end{IEEEbiography}

\begin{IEEEbiographynophoto}{Coauthor}
Same again for the co-author, but without photo
\end{IEEEbiographynophoto}

\end{document}